\documentclass[epj]{svjour}
\usepackage{graphics}
\begin{document}
\title{Doppler-free spectroscopy in driven three-level
systems}
\author{Umakant D. Rapol \and Vasant Natarajan
\thanks{Electronic mail: vasant@physics.iisc.ernet.in}}
\institute{Department of Physics, Indian Institute of Science,
Bangalore 560 012, INDIA}
\date{Received: date / Revised version: date}
\abstract{We demonstrate two techniques for studying the
features of three-level systems driven by two lasers
(called control and probe), when the transitions are
Doppler broadened as in room-temperature vapor. For
$\Lambda$-type systems, the probe laser is split to produce
a counter-propagating pump beam that saturates the
transition for the zero-velocity atoms. Probe transmission
then shows Doppler-free peaks which can even have
sub-natural linewidth. For V-type systems, the transmission
of the control beam is detected as the probe laser is
scanned. The signal shows Doppler-free peaks when the probe
laser is resonant with transitions for the zero-velocity
group. Both techniques greatly simplify the study of
three-level systems since theoretical predictions can be
directly compared without complications from Doppler
broadening and the presence of multiple hyperfine levels in
the spectrum.
\PACS{
      {42.50.Gy}{Effects of atomic coherence on propagation,
 absorption, and amplification of light}  \and
      {42.50.-p}{Quantum optics}
     } 
} 
\maketitle

\section{Introduction}
There have been several recent theoretical and
experimental studies where control lasers have been used
in three-level systems to modify the absorption
properties of a weak probe laser
\cite{NSO90,ZLN95,VAR96,ZHW96}. For example, in
electromagnetically induced transparency (EIT), an
initially absorbing medium is made transparent to a probe
beam when a strong control laser is turned on
\cite{BIH91,HAR97}. EIT techniques have several practical
applications in probe amplification \cite{MEA00}, lasing
without inversion \cite{ZLN95} and suppression of
spontaneous emission \cite{GZM91,ZNS95,AGA96}.
Experimental observations of EIT have been facilitated by
the advent of low-cost tunable diode lasers which can be
used to access transitions in alkali atoms such as Rb and
Cs. Alkali atoms have convenient energy levels with
strong oscillator strengths which form almost ideal
three-level systems. However, spectroscopy in
room-temperature vapor of these atoms is often limited by
Doppler broadening, which is typically a factor of 100
larger than the natural linewidth. Therefore, in many
experiments, large laser intensities are needed to
overcome this broadening and observe the predicted
effects \cite{ZHW96}, since the scale for the Rabi
frequency of the control laser is set by the width of the
transition. Moreover, the Doppler-broadened spectrum
contains several closely-spaced hyperfine levels, so that
the system is not a simple three-level system but a
multi-level system. In such situations, theoretical
calculations based on the three-level model are only
approximately valid.

In this paper, we demonstrate two techniques to overcome
the Doppler broadening in room-temperature vapor and make
the system a true three-level system. The first technique
applies to $\Lambda$-type systems, where the control and
probe lasers couple different ground levels to the same
excited level. The second technique applies to V-type
systems, where the two lasers couple the same ground level
to different excited levels. We demonstrate these
techniques in a room-temperature vapor of Rb atoms, with
two diode lasers tuned to different hyperfine transitions
in the $D_2$ line. We obtain Doppler-free spectra and
observe predicted features such as sub-natural linewidths
for the dressed states created by the control laser
\cite{RWN03}. The separation and linewidth of the states
depends on control-laser intensity as predicted from a
simple model of a driven two-level system. In addition, the
creation of resonances with sub-natural linewidth has
applications in precision spectroscopy \cite{RAN02b},
better stabilization of lasers to atomic transitions, and
attaining sub-Doppler temperatures in laser cooling of
atoms and ions \cite{RLM00}.

\section{Experimental details}
The experimental set up, shown schematically in Fig.\ 1,
consists of a probe beam and a control beam derived from
two frequency-stabilized diode laser systems operating near
the $D_2$ line in $^{87}$Rb ($5S_{1/2} \leftrightarrow
5P_{3/2}$ transition at 780 nm). The linewidth of the
lasers after stabilization is less than 1 MHz. The probe
beam has $1/e^2$ diameter of 3 mm while the control beam is
slightly larger with diameter of 4 mm. The two beams are
mixed in a beamsplitter and co-propagate through a
room-temperature vapor cell containing Rb. The absorption
through the cell is about 25\%. The two beams have
identical polarization and the angle between them is about
17 mrad. For some experiments, a counter-propagating pump
beam is generated from the probe laser using a
beamsplitter.

The $D_2$ line in $^{87}$Rb has two hyperfine levels in
the ground state ($F=1,2$) and four hyperfine levels in
the excited state ($F'=0,1,2,3$). Thus, as shown in Fig.\
2, it can be used as either a $\Lambda$ system or a V
system, depending on which levels are coupled by the
lasers. For the $\Lambda$ system, the same excited level
($F'=2$) is coupled to two ground levels: to the $F=1$
level by the control laser and to the $F=2$ level by the
probe laser. For the V system, the same ground level
($F=2$) is coupled to two excited levels: to the $F'=3$
level by the control laser and the $F'=1$ (or 2) level by
the probe laser. The control laser has Rabi frequency of
$\Omega_R$ and detuning from resonance of $\Delta_c$. The
weak probe laser has detuning $\Delta$. The spontaneous
decay rate from the excited levels is $\Gamma$, which is
$2\pi \times 6.1$ MHz in Rb.

\section{$\Lambda$-system}
We first consider the $\Lambda$ system. The absorption of
the weak probe is proportional to Im$(\rho_{13})$, where
$\rho_{13}$ is the induced polarization on the $\left| 1
\right> \leftrightarrow \left| 3 \right>$ transition
coupled by the probe laser. From the density-matrix
equations, the steady-state value of $\rho_{13}$ is given
by \cite{VAR96}:
\begin{equation}
\rho_{13} = \frac{(\Omega_p/2)(\Delta - \Delta_c)}{\left| \Omega_R/2
\right|^2 -
i(\Gamma - i\Delta) (\Delta - \Delta_c)},
\end{equation}
where $\Omega_p$ is the Rabi frequency of the (weak) probe beam.
The pole structure of the above equation shows that
probe absorption has two peaks at the following
detunings:
\begin{equation}
\Delta_{\pm} = \frac{\Delta_c}{2} \pm
\frac{1}{2}\sqrt{\Delta_c^2 + \Omega_R^2},
\end{equation}
with the corresponding linewidths:
\begin{equation}
\Gamma_{\pm} = \frac{\Gamma}{2}
\left( 1 \mp \frac{\Delta_c}{\sqrt{\Delta_c^2 +
\Omega_R^2}} \right) .
\end{equation}
The two peaks are the well-known Autler-Townes doublet
appearing at the location of the two dressed states
created by the control laser \cite{COR77}. Due to
coherence between the dressed states, the peaks have
asymmetric linewidths for non-zero detuning of the
control laser, in such a way that the sum of the two
linewidths is equal to the unperturbed linewidth,
$\Gamma$.

The above analysis shows how three-level systems are
useful in many applications. For example, probe
absorption at line center ($\Delta=0$) is strongly
suppressed in the presence of a resonant control laser
because the dressed states created by the control laser
are shifted by the Rabi frequency. This is the basis for
EIT experiments. Similarly, it is seen from Eq.\ 3 that
the linewidth $\Gamma_+$ of the second dressed state can
be much below the natural linewidth when the
control-laser detuning is large.

However, observing such effects in room-temperature vapor
is complicated by effects due to Doppler broadening. The
above expressions are valid for a stationary atom; in
room-temperature vapor they have to be averaged over the
Maxwell-Boltzmann velocity distribution of the atoms
taking into account the different frequencies seen by
each atom in its velocity frame. The effect of this
averaging is that the location of the peaks given in Eq.\
2 does not change, but the linewidth of the peaks changes
to \cite{VAR96}:
\begin{equation}
\Gamma_{\pm} = \frac{\Gamma + D}{2}
\left( 1 \mp \frac{\Delta_c}{\sqrt{\Delta_c^2 +
\Omega_R^2}} \right) ,
\end{equation}
where $D$ is the Doppler width, which is 560 MHz for
room-temperature Rb atoms. Thus, in EIT experiments, the
Rabi frequency of the control laser has to be very large
to see significant reduction in absorption at line
center. Another complication due to the appearance of the
Doppler width is that the different excited-state
hyperfine levels, spaced a few 100 MHz apart, all lie
within the Doppler-broadened profile. Any comparison with
the predictions of the above equations is difficult
because the control laser couples to several hyperfine
levels, but with different detunings. The simple model of
a two-level system driven by the control laser is no
longer valid.

We have solved this problem for the $\Lambda$ system in the
following manner. A part of the probe laser is split off
and sent through the cell so that it is counter-propagating
with respect to the probe and control beams. The intensity
of this pump beam is chosen to be about 5 times higher than
the probe. In this configuration, the zero-velocity group
of atoms preferentially absorbs from the pump and the probe
gets transmitted. This is a standard technique used in
Doppler-free saturated-absorption spectroscopy
\cite{DEM82}, which we have adapted to the three-level
case. Note that the pump beam is still very weak compared
to the control and any coherent driving from this beam can
be neglected as in the case of the probe beam \cite{AGA02}.

In Fig.\ 3a, we show the probe spectrum as the laser is
scanned across all the $F=2 \leftrightarrow F'$
transitions. The control laser is locked to the $F=1
\rightarrow F'=2$ transition with a detuning of $\Delta_c =
-11.5$ MHz. With the control beam off, the spectrum shows
the typical Doppler-broadened profile (with a linewidth of
560 MHz), and ``Doppler-free'' peaks at the location of the
hyperfine transitions due to saturation by the pump beam.
The words ``Doppler-free'' are set in quotes because the
linewidth of the hyperfine peaks is usually larger than the
natural linewidth. The primary cause for this is power
broadening from the pump laser and a small misalignment
angle between the counter-propagating beams. Indeed, we
have studied the variation of the linewidth as a function
of pump power. With near-perfect alignment of the beams and
with a magnetic shield \cite{mag} around the cell, the
linewidth extrapolated to zero power is only 6.5 MHz, close
to the natural linewidth of 6.1 MHz. This shows that
collisional broadening in the vapor cell is negligible. The
effect of stray magnetic fields in the vicinity of the cell
is to increase the linewidth by 15--20\%, which we have
verified by measuring the linewidth with and without the
magnetic shield.

The near-perfect alignment of the pump-probe beams is
possible only by mixing and separating the beams using
polarizing-beamsplitter cubes. This requires the use of
orthogonal polarization for the two beams. For the
experiments reported here, we have not used this technique
but rather used a small angle (of $\sim$17 mrad) between
the beams. There are two reasons for this. The first is
that we require the beams to have the same polarization.
Otherwise, the different magnetic sublevels are driven
differently, and the analysis becomes complicated. The
second reason is that the separation of
orthogonal-polarization beams using a polarizing
beamsplitter is not perfect. Any leakage of one beam into
the other makes the interpretation of the detected signal
questionable. Thus, under typical conditions, the observed
linewidth has three broadening mechanisms: power broadening
due to the pump beam, which increases the linewidth from
6.1 MHz to about 20 MHz; misalignment by 17 mrad, which
increases it further to 25 MHz, and residual magnetic
fields, which increases it to 30 MHz. This accounts for the
linewidth of the hyperfine peaks in Fig.\ 3a, but note that
this is still 20 times smaller than the Doppler width and
is thus essentially Doppler free.

As seen in the bottom trace in Fig.\ 3a, when the control
laser is turned on, the $F'=2$ peak splits into two as
predicted by the earlier analysis of the three-level
system. The two peaks are clearly resolved even though
their separation is only of order 20 MHz. As shown in the
inset, the peaks also have asymmetric linewidths. This is
because of the non-zero detuning of the control laser
($\Delta_c = -11.5$ MHz), which results in unequal
linewidths from Eq.\ 3.

We have thus achieved a situation where the Doppler
broadening is virtually eliminated and the three-level
assumption is valid. Therefore the predictions of Eqs.\ 2
and 3 can be applied without worrying about the presence of
other hyperfine levels. To demonstrate this, we have
studied the probe-absorption spectra as a function of
control-laser power and detuning. As shown in Fig.\ 3b, the
separation of the peaks \cite{foot1} increases with power as
expected from Eq.\ 2. We have plotted the separation as a
function of control-laser power and not Rabi frequency
since the power is what is measured in the laboratory. The
relation between the power and the Rabi frequency depends
on factors such as the intensity profile across the control
beam, overlap between the beams, losses at the cell
surfaces, absorption through the cell, and so on. It is
difficult to experimentally determine these factors with
any degree of certainty, therefore we have left this factor
as an overall fit parameter in obtaining the solid line in
Fig.\ 3b. It is important to note that this fit parameter
does not determine the trend in the data, and changes in
its value would cause the solid line to move up or down
without changing its shape. The best fit shown in the
figure is obtained with a parameter that assumes that the
control-laser power is spread uniformly over a circle of
diameter 4.6 mm, which is reasonable given that the
measured size of the Gaussian beam is 4 mm.

The linewidth of the smaller peak also follows the trend
predicted by Eq.\ 3, as shown in Fig.\ 3c. There are no
additional fit parameters to obtain the solid line except
that we have to increase the unperturbed linewidth in Eq.\
3 from the natural linewidth of 6.1 MHz to 30 MHz. As
explained earlier, this is the linewidth obtained for the
hyperfine peak without the control laser. The two points at
low power lie slightly above the curve, possibly because in
these cases it is hard to determine the linewidth
accurately since the lineshape is distorted by the
underlying Doppler profile and peak pulling by the other
dressed state. However, despite these effects and the
somewhat large unperturbed linewidth, the linewidth of the
smaller peak is below the natural linewidth for small
control powers. We have recently demonstrated this as a
technique to achieve sub-natural resolution in room
temperature vapor \cite{RWN03}. We have also combined the
narrow linewidth and the variation with detuning to make
precise measurements of hyperfine intervals. We have used
this technique to measure a hyperfine interval in Rb with
an accuracy of 44 kHz \cite{RAN02b}.

\section{V-system}
We now turn to the V-system. The theoretical analysis
proceeds along similar lines as for the $\Lambda$-system
\cite{MEA00}. In room temperature vapor, the location of
the two Autler-Townes peaks is the same as before (Eq.\ 2).
Similarly, the linewidths of the dressed states (given by
Eq.\ 3) are increased by the Doppler width after thermal
averaging. In this case, we obtain Doppler-free spectra not
by using a counter-propagating pump beam but by detecting
the transmission of the control laser through the vapor
cell. The control laser is locked to $F=2 \leftrightarrow
F'=3$ transition while the probe laser is scanned across
all the $F=2 \leftrightarrow F'$ transitions. Since the
control laser is locked, it is only the zero-velocity atoms
that absorb from this laser. When the probe laser is also
resonant with a transition starting from the same
zero-velocity atoms, the absorption from the control laser
is slightly reduced and the transmitted signal shows a
narrow peak. Thus, as the probe laser is scanned,
Doppler-free peaks appear in the transmission of the
control-laser at the exact locations where the probe is
absorbed.

The resultant spectra are shown in Fig.\ 4. The
experimental parameters are similar to the case of the
$\Lambda$ system, except that the beam sizes are reduced by
a factor of 3, and there is no counter-propagating pump
beam. Under these conditions, the probe spectrum shows the
expected Doppler broadening with some increased
transmission at line center for the $F'=1$, 2 and 3
transitions. However, there is no apparent signature of the
dressed states created by the control laser. By contrast,
the control transmission clearly shows the Autler-Townes
doublets near the $F'=1$ and 2 levels. It is interesting to
note the increased transmission of the probe laser at line
centers. In the case of the $F'=1$ and 2 levels, the
increased transmission is precisely due to EIT because the
location of the dressed states is shifted away from the
line center. Indeed, at higher power (dotted line in
figure), the transparency approaches 100\%. In the case of
the $F'=3$ level, both probe and control lasers couple the
same levels and the increased transmission of the probe is
due to saturating effects from the much stronger control
laser.

One of the disadvantages of the pump-probe technique used
for the $\Lambda$ system is that the Doppler-free peaks
appear on top of a broad Doppler profile, as seen in Fig.\
3a. However, in the V system, our technique of measuring
the control-laser transmission results in a constant
background signal corresponding to absorption by the
zero-velocity atoms. The signal changes only when the
zero-velocity atoms start absorbing from the probe.
Therefore, the Doppler-free peaks appear on a flat
background. This is shown in Fig.\ 5a, which is a waterfall
plot of control transmission spectra for different values
of control power. The $F'=1$ and 2 peaks split into
Autler-Townes doublets, with the separation of the doublets
increasing with power. At very low powers (below 0.5 mW)
the splitting is smaller than the unperturbed linewidth,
and the doublet peaks are not well resolved. One other
feature of the spectrum is that the doublet peaks in the
$F'=2$ level are not of equal height. This kind of
asymmetry has also been observed in the case of
Doppler-broadened doublets \cite{ZHW96}.

As in the case of the $\Lambda$ system, the Doppler-free
spectra allow the predictions of the theoretical analysis
to be applied with confidence. This is seen in Fig.\ 5b,
where we plot the separation of the peaks in the
Autler-Townes doublet as a function of control-laser power.
The solid line is the variation predicted by Eq.\ 2. As
before, the relation between the power and the Rabi
frequency is left as a fit parameter. With the best fit,
the agreement with the observed separation is quite good,
indicating that the three-level model is adequate to
explain the data.

The analysis of the linewidth of the two peaks is slightly
different from the case of the $\Lambda$ system. Since the
control-laser detuning is zero, Eq.\ 3 predicts that the
linewidth should remain constant at $\Gamma/2$. From Fig.\
5c, we see that the measured linewidth is about 12 MHz
(corresponding to a total linewidth of 24 MHz) and shows a
gradual increase as the power is increased. If we assume an
unperturbed linewidth of 7 MHz, then it increases to 14 MHz
due to the misalignment and further to 17 MHz due to stray
magnetic fields, which is still lower than the measured
value. We attribute the remaining linewidth to power
broadening by the control laser. In the case of the
$\Lambda$ system, the control laser is detuned from
resonance and its influence on the linewidth is negligible,
therefore the only power broadening comes from the pump
beam.

The power-broadened linewidth varies as $\Gamma
\sqrt{1+s}$, where $s$ is the saturation parameter.
However, at the control-laser powers used in the
experiment, the power-broadened linewidth is 40 MHz, which
is much larger than the observed linewidth of 24 MHz. This
is because the control laser is primarily driving the atoms
coherently, and power broadening is a smaller effect. To
account for this reduction, we have fitted the linewidth to
$\Gamma \sqrt{1+ks}$, with $k$ as a fit parameter. In Fig.\
5c, the solid line is obtained with a value of $k=0.16$, or
16\%. The curve fits the data reasonably well and explains
why the linewidth increases with control power. However, it
is important to note that the measured linewidth is still
much smaller than the Doppler width, as in the case of the
$\Lambda$ system.

\section{Conclusions}
We have thus demonstrated techniques to overcome Doppler
broadening when studying the properties of three-level
systems driven by two lasers. The $\Lambda$ system requires
the use of a counter-propagating pump beam to saturate the
transition for the zero-velocity atoms. In the V system,
the control laser transmission is detected as the probe
laser is scanned. The transmitted signal shows Doppler-free
peaks when the probe is absorbed by the zero-velocity
atoms. In the $\Lambda$ system, probe absorption is to the
two dressed states in the upper level created by the
control laser. On the other hand, in the V system, probe
absorption is from the two dressed states created in the
ground state. Probably because of this difference, each
technique works only in its respective system and not in
the other system. But the techniques allow the predictions
of simple theoretical models to be applied directly since
there are no complications arising from Doppler broadening
or the presence of multiple levels in the Doppler-broadened
spectrum. We expect these techniques to find widespread use
in the study of driven three-level systems. We have already
demonstrated the application of these ideas for
high-resolution hyperfine measurements in $\Lambda$-systems
\cite{RAN02b}. In future, we plan to apply such ideas to
precision spectroscopy in V-systems also.

\section{Acknowledgments}

We thank Anusha Krishna, S. Shivakumar, Rajat Thomas, and
Pierre Mangeol for help with the experiments. This work was
supported by the Department of Science and Technology,
Government of India.

\begin{figure}
\resizebox{1\columnwidth}{!}{\includegraphics{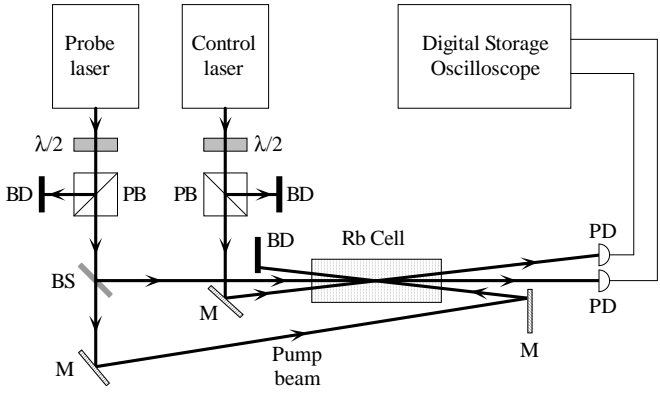} }
\caption{ Schematic of the experiment. The probe and
control beams are derived from frequency-stabilized,
tunable diode laser systems. The angle between the beams
has been exaggerated for clarity; in reality it is close to
0, ensuring good overlap of the beams in the Rb cell. }
\label{fig1}
\end{figure}

\begin{figure}
\resizebox{1\columnwidth}{!}{\includegraphics{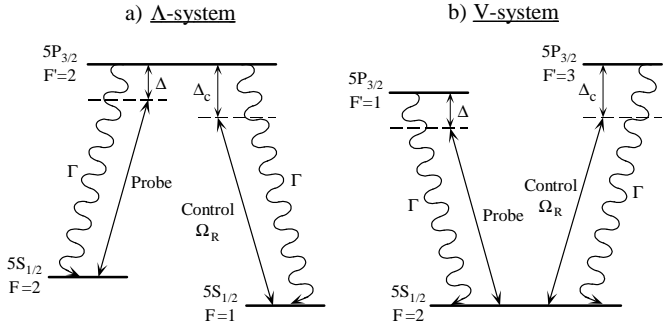} }
\caption{ Three-level systems in Rb. Rb can be used to form
either a $\Lambda$ system or a V system, depending on which
hyperfine levels are coupled by the two lasers. The control
laser has detuning $\Delta_c$ and Rabi frequency
$\Omega_R$. The probe laser has detuning $\Delta$.}
\label{fig2}
\end{figure}

\begin{figure}
\resizebox{1\columnwidth}{!}{\includegraphics{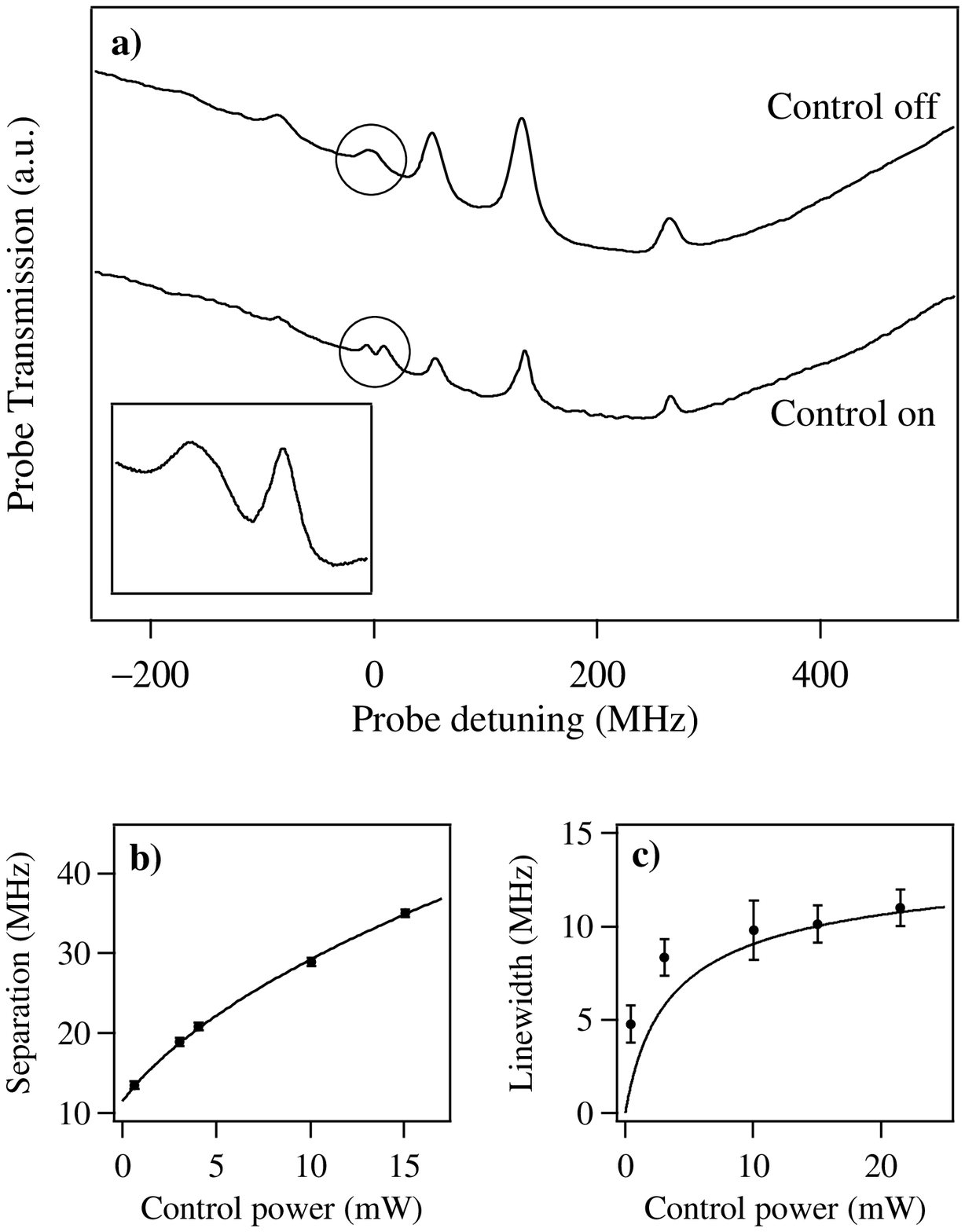} }
\caption{ Doppler-free spectroscopy in $\Lambda$ system. In
a), we show the transmission of the probe laser as it is
scanned across all the $F=2 \leftrightarrow F'$
transitions. The upper trace is with the control beam
turned off and shows the usual saturated absorption peaks.
The lower trace is with the control beam turned on, and
shows the $F'=2$ peak splitting into an Autler-Townes
doublet. The control laser has detuning of $-11.5$ MHz. The
inset is a close-up of the doublet showing the asymmetric
linewidths of the two peaks. In b) and c), we plot the
separation of the peaks and the linewidth of the smaller
peak as a function of control-laser power. The solid lines
are the expected variation from Eqs.\ 2 and 3, as explained
in the text.} \label{fig3}
\end{figure}

\begin{figure}
\resizebox{1\columnwidth}{!}{\includegraphics{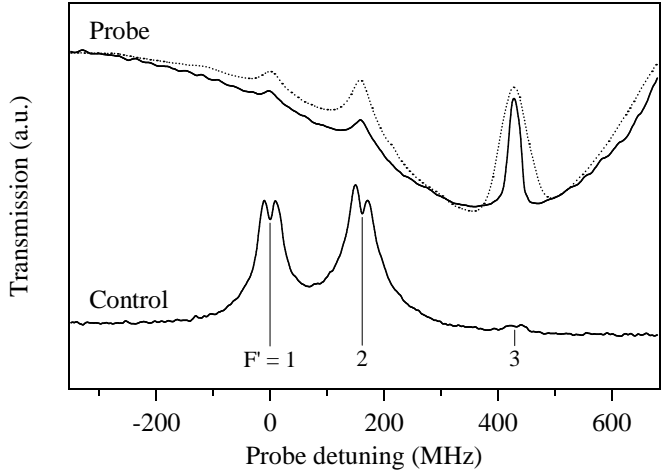} }
\caption{ Spectroscopy in V system. The upper trace is the
transmission of the probe laser as it is scanned across all
the $F=2 \leftrightarrow F'$ transitions. The spectrum
shows the expected Doppler broadening. The peaks at $F'=1$
and 2 are due to EIT effects from a resonant control laser.
The transparency increases to nearly 100\% as the
control-laser power is increased from 0.8 mW (solid line)
to 7 mW (dotted line). The peak at $F'=3$ is due to
saturation by the control laser. The lower trace is the
transmission of the control laser. The Autler-Townes
doublets near the $F'=1$ and 2 levels are clearly resolved
with Doppler-free linewidths.} \label{fig4}
\end{figure}

\begin{figure}
\resizebox{1\columnwidth}{!}{\includegraphics{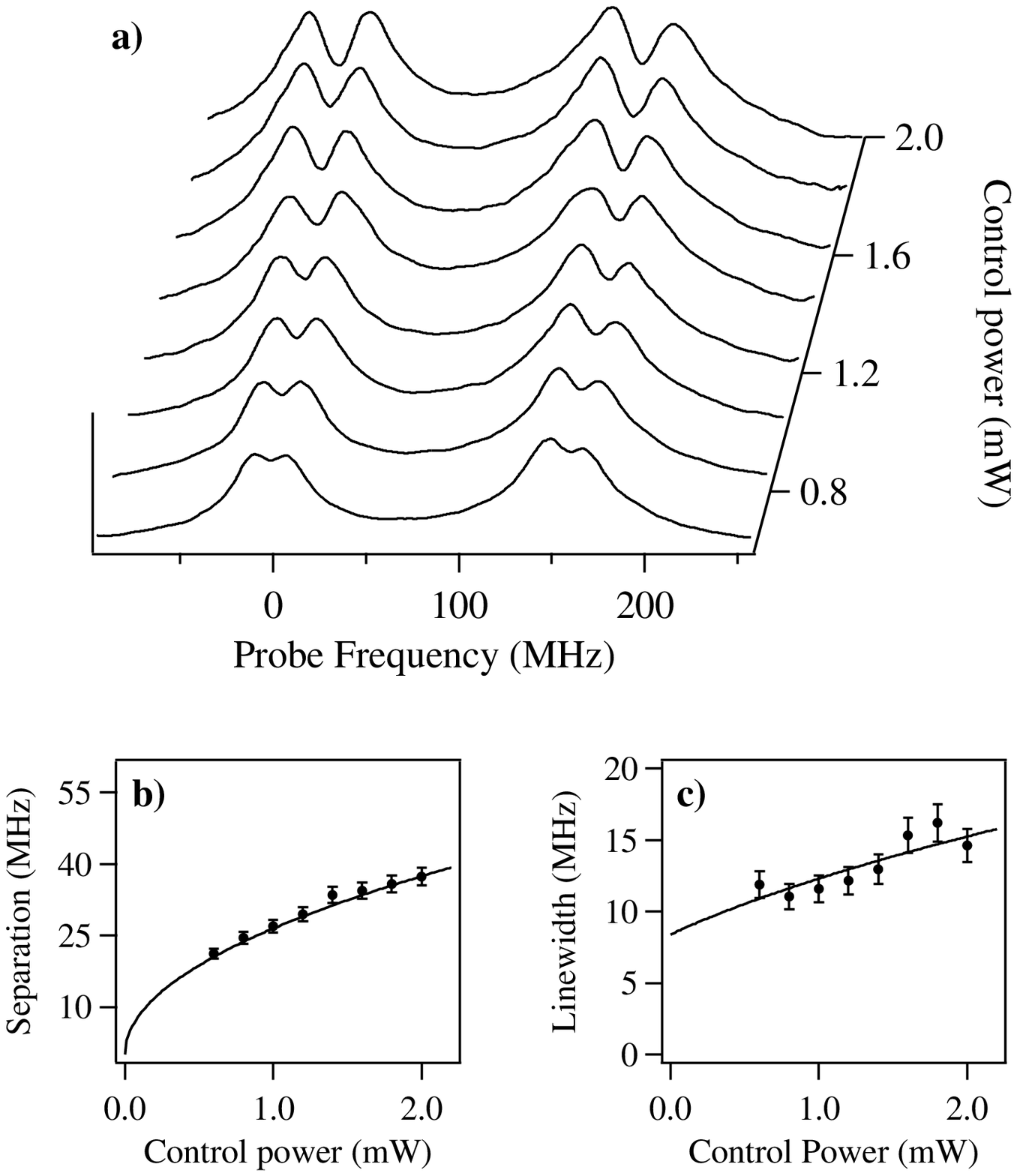} }
\caption{ In a), we show a waterfall plot of the
transmission of the control laser as a function of
probe-laser detuning, for different values of control-laser
power. The control laser has zero detuning. As the control
power is increased, the signal near the $F'=1$ and 2 levels
shows the Autler-Townes doublets. In b), we plot the
separation of the peaks as a function of power. The solid
line is the variation expected from Eq.\ 2 in the text. In
c), we show the linewidth of one of the peaks as a function
of control power. The solid line is the increase due to
power broadening, as explained in the text.} \label{fig5}
\end{figure}


\begin{thebibliography}{}

\bibitem{NSO90}
L.~M. Narducci {\it et~al.}, Phys. Rev. A. {\bf 42},
1630  (1990).

\bibitem{ZLN95}
A.~S. Zibrov {\it et~al.}, Phys. Rev. Lett. {\bf 75},
1499  (1995).

\bibitem{VAR96}
G. Vemuri, G.~S. Agarwal, and B.~D. Nageswara Rao,
Phys. Rev. A. {\bf 53},  2842  (1996).

\bibitem{ZHW96}
Y. Zhu and T.~N. Wasserlauf, Phys. Rev. A. {\bf 54},
3653  (1996).

\bibitem{BIH91}
K.-J. Boller, A. Imamo\u{g}lu, and S.~E. Harris, Phys.
Rev. Lett. {\bf 66},  2593
  (1991).

\bibitem{HAR97}
S.~E. Harris, Phys. Today {\bf 50},  36  (1997).

\bibitem{MEA00}
S. Menon and G.~S. Agarwal, Phys. Rev. A. {\bf 61},
013807  (2000).

\bibitem{GZM91}
D.~J. Gauthier, Y. Zhu, and T.~W. Mossberg, Phys. Rev.
Lett. {\bf 66},  2460
  (1991).

\bibitem{ZNS95}
S.-Y. Zhu, L.~M. Narducci, and M.~O. Scully, Phys. Rev.
A. {\bf 52},  4791
  (1995).

\bibitem{AGA96}
G.~S. Agarwal, Phys. Rev. A. {\bf 54},  R3734  (1996).

\bibitem{RWN03}
U.~D. Rapol, A. Wasan, and V. Natarajan, Phys. Rev. A. {\bf
67}, 053802  (2003).

\bibitem{RAN02b}
U.~D. Rapol and V. Natarajan, Europhys. Lett. {\bf 60},
195  (2002).

\bibitem{RLM00}
C.~F. Roos {\it et~al.}, Phys. Rev. Lett. {\bf 85},  5547
(2000).

\bibitem{COR77}
C. Cohen-Tannoudji and S. Reynaud, J. Phys. B. {\bf 10},
365  (1977).

\bibitem{DEM82}
W. Demtr\"oder, {\em Laser Spectroscopy}
(Springer-Verlag, Berlin, 1982).

\bibitem{AGA02}
G.~S. Agarwal, private communication.

\bibitem{mag}
Conetic AA Alloy, Magnetic Shield Division, Perfection Mica
Co., U.S.A.

\bibitem{foot1}
The separation and linewidth of the two peaks are obtained
by fitting to the spectrum with an underlying Doppler
profile.

\end{thebibliography}
\end{document}